\documentclass[a4paper,11pt]{article}

\usepackage{subfig}
\usepackage{pgfplots}
\usepackage{pgfplotstable}
\usepackage{mathtools}
\usepackage{multicol}
\usepackage{comment}
\usepackage{booktabs}
\pgfplotsset{compat=1.5}
\usepackage{amssymb}
\usepackage{url}
\usepackage{bm}

\usepackage{pdfsync}
\usepackage{float}
\usepackage{tabularx}
\usepackage{enumerate}
\usepackage{array}
\usepackage{xspace}

\usepackage{authblk}

\newcommand{\mupar}{\ensuremath{\boldsymbol{\mu}}}

\newcommand{\zetapar}{\ensuremath{\boldsymbol{\zeta}}}

\begin{document}

\title{Efficient Reduction in Shape Parameter Space Dimension for Ship Propeller Blade Design}

\author[1]{Andrea~Mola\footnote{andrea.mola@sissa.it}}
\author[1]{Marco~Tezzele\footnote{marco.tezzele@sissa.it}}
\author[1]{Mahmoud~Gadalla\footnote{mgadalla@sissa.it}}
\author[2]{Federica~Valdenazzi\footnote{federica.valdenazzi@cetena.it}}
\author[2]{Davide~Grassi\footnote{davide.grassi@cetena.it}}
\author[3]{Roberta~Padovan\footnote{roberta.padovan@cetena.it}}
\author[1]{Gianluigi~Rozza\footnote{gianluigi.rozza@sissa.it}}

\affil[1]{Mathematics Area, mathLab, SISSA, International School of
  Advanced Studies, via Bonomea 265, I-34136 Trieste, Italy}
\affil[2]{CETENA S.p.A., via Ippolito D'Aste 5, I-16121 Genova, Italy}
\affil[3]{CETENA S.p.A., Branch Office Trieste Passeggio S. Andrea, 6/A - 34100 Trieste, Italy}

\maketitle

\begin{abstract}
In this work, we present the results of a ship propeller design optimization campaign carried out in the framework of the research
project PRELICA, funded by the Friuli Venezia Giulia regional government. The main idea of this work is to operate on a multidisciplinary
level to identify propeller shapes that lead to reduced tip vortex-induced pressure and increased efficiency without altering the thrust.
First, a specific tool for the bottom-up construction of parameterized propeller blade geometries
has been developed. The algorithm proposed operates with a user defined number of arbitrary shaped or NACA airfoil sections, and
employs arbitrary degree NURBS to represent the chord, pitch, skew and rake distribution as a function of the blade radial
coordinate. The control points of such curves have been modified to generate, in a fully automated way, a family of blade geometries
depending on as many as 20 shape parameters. Such geometries have then been used to carry out potential flow simulations with the
Boundary Element Method based software PROCAL. Given the high number of parameters considered, such a preliminary
stage allowed for a fast evaluation of the performance of several hundreds of shapes. In addition, the data obtained from the
potential flow simulation allowed for the application of a parameter space reduction methodology based on active subspaces (AS)
property, which suggested that the main propeller performance indices are, at a first but rather accurate
approximation, only depending on a single parameter which is a linear combination of all the original geometric ones. AS analysis has
also been
used to carry out a constrained optimization exploiting response surface method in the reduced parameter space, and a sensitivity
analysis based on such surrogate model. The few selected shapes were finally used to set up high fidelity RANS simulations and
select an optimal shape.
\end{abstract}


\section{Introduction}
\label{sec:intro}
In several fields of engineering, virtual prototyping simulations results depend on a wide range of different design
parameters. When the number of such input parameters becomes too large, the problem of finding their combination resulting
in the optimal solution can be easily affected by the curse of dimensionality. Depending on the computational cost of the
single simulations, even with a relatively small parameter space dimension, a full optimization campaign could require months
to be completed. Thus, reducing the dimension of such space is crucial to allow for quality optimization in engineering design
processes.

In recent years, several interesting applications of shape parameter reduction techniques have been been documented in the literature.
Among them, we mention \cite{d2017nonlinear,d2018deep}, in which the authors apply both nonlinear extensions of the Principal
Component Analysis (PCA)~\cite{Schlkopf1998NonlinearCA} and methods based on Artificial Neural Networks (ANN)~\cite{HintonSalakhutdinov2006b}
to approximate in low dimensional spaces the parametric deformation of ship hulls. A common feature of such works,
is that they act in an \emph{offline} fashion, since they solely operate on the relationship between shape parameters and
hull geometry, rather than on the one between shape parameters and simulations output. This leads to the advantage that less
simulations are required in the \emph{online} optimization phase. In this work, we make instead use of an analysis based on
the Active Subspaces (AS) property~\cite{constantine2015active} to obtain parameter space reduction in the framework of
a ship propeller shape optimization campaign. A main trait of the present analysis is that, differently from the ones described,
it is carried out in the online phase of
the optimization so as to construct a reduced parameter space to approximate the relationship between the simulations
output and the parameters. Although this might lead to increased computational cost, the analysis has the fruitful benefit of
identifying which of the original parameters bear a higher influence on the physical output. Such information can of course
lead the work of design engineers. In addition, to mitigate the disadvantage of possibly high computational cost associated
to the high number of simulations required for the analysis, in this work we made use of the potential flow solver
PROCAL \cite{vazPROCAL}, which despite its low computational cost, is able to provide accurate predictions of the fluid dynamic outputs
of interest. Moreover, we also explore the use of AS for constrained optimization exploiting response surface method in the reduced
parameter space, to identify propeller shapes with increased hydroacoustic performance (i.e.: reduced tip vortex-induced maximum pressure)
without thrust reductions. The most promising  shapes are the only ones tested with the high-fidelity RANS solver, with
considerable reduction of the whole optimization campaign.

\section{Blade reconstruction and morphing}
\label{sec:blade}

A very important ingredient of the multidisciplinary propeller optimization methodology here described is represented by an
efficient shape parameterization tool. In fact, as well known, optimization algorithms are mathematical tools which operate on
numerical variables, identifying the input parameters combination which maximizes or minimizes the output values of a specific
model or system. In such framework, optimization algorithms cannot be used to find shapes of optimal performance, unless a shape
parameterization strategy is devised to associate each possible shape modification with numbers characterizing the points
in the parameter space. Such numbers are the input used to feed the optimization algorithm.
Thus, the main task of shape parameterization is that of creating a --- possibly --- one-to-one correspondence between
propeller shapes and sample points in the parameter space. There are several multi-purpose parameterization methodologies available
in the literature, which are designed to deform bodies of arbitrary shapes. Such algorithms, among which we mention
Free Form Deformation (FFD)~\cite{rozza2013free} and Radial Basis Functions (RBF), are implemented in open source software
libraries and packages~\cite{pygem,salmoiraghi2016advances,tezzele2018ecmi} which could be in principle readily downloaded and employed.  
Unfortunately, in their original formulation such multi-purpose deformation strategies are not suitable for a highly
engineered shape as a ship propeller. Among other things, their application would in fact result in altering in an
undesired way the specific airfoils selected by the engineers at each blade section for their well assessed hydrodynamic performance.
Rather than tweaking FFD or RBF to account for constraints on the shape deformations generated, we decided to exploit the procedure
used by the engineers for the bottom-up generation of 3D propeller geometries.

\subsection{Bottom-up blade construction of parameterized propeller}

\begin{figure}[hbt!]
\centering
\subfloat[][]{\includegraphics[width=0.52\textwidth]{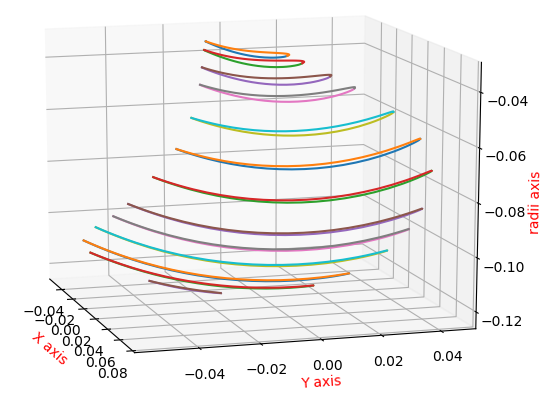}} \hspace{3mm}
\subfloat[][]{\includegraphics[width=0.41\textwidth]{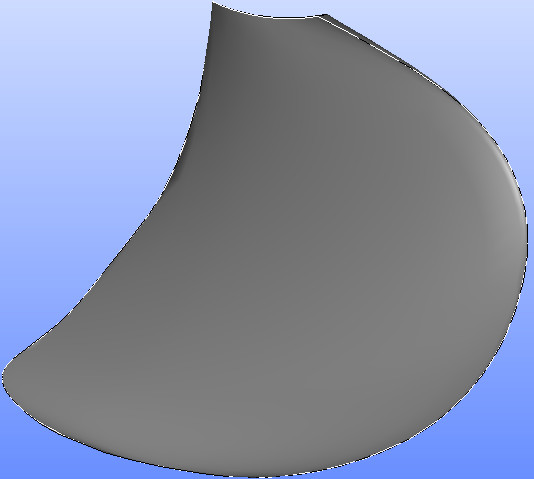}}
\caption{PPTC blade bottom-up construction with BladeX: (a) Cylindrical blade sections in their final three dimensional position.
         (b) The generated CAD surface (saved in \emph{iges} file format).} \label{iges_PPTC}
\end{figure}

A 3D propeller blade is generated (see for
instance \cite{carlton1994marine}) as the surface passing through a set of sectional airfoil shapes, which are originally
specified in a 2D space and are successively located in the 3D space according to a set of transformations which vary
along with the radial coordinate of each section. Such transformations include scaling, translations and rotations to
obtain the blade with the desired radial distribution of airfoil section chord length, rake and skew displacements, and
pitch angle respectively. Such standard propeller blade design procedure has been implemented in the open source
python package BladeX ~\cite{gadalla19bladex}. As illustrated in Figure~\ref{iges_PPTC}, after the coordinates of blade airfoil
sections and radial distribution curves are read from external files, the airfoil sections are placed in the correct three dimensional
position and the CAD surface passing through the sections is generated and exported in iges format.

\begin{figure}[hbt!]
\centering
\includegraphics[width=0.9\textwidth]{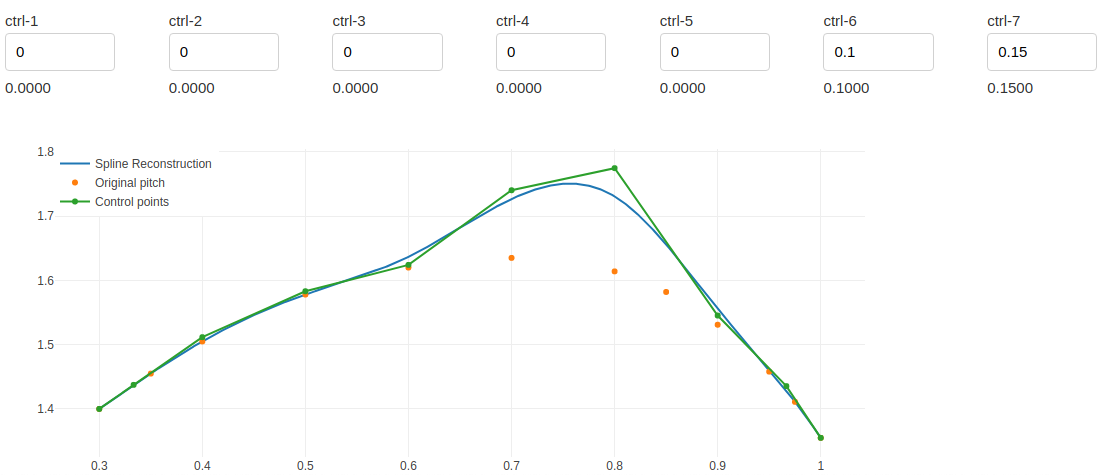}
\caption{A sketch of the PPTC blade pitch radial distribution curve modification carried out with BladeX. The plot shows
         the original blade points (yellow dots), and the corresponding splines reconstruction (blue continuous line)
         with its control polygon (green dots and line). The example shows how control points 6 and 7 are modified to
         alter the pitch curve retaining smoothness.} \label{bladex_PPTC}
\end{figure}

In the framework of the described blade construction procedure, BladeX allows for reconstructing with user specified degree splines,
the radial distribution curves for chord length, pitch angle, skew and rake displacements. By means of constrained least squares
minimization, the algorithm will in fact identify the spline control points position minimizing the distance between the original
curves and their splines counterparts. A further method has been added to allow for spline reconstruction of the radial distribution
of the sectional airfoils maximum camber deflection. 
Once chord, pitch, skew, rake and camber radial distributions have been reconstructed by means of splines, the user introduces a set of
splines control points displacements to alter the blade characteristic curves and ultimately its shape. Thus, a parametererized blade
geometry can be generated through variations of the position of an arbitrary number of the control points associated with the spline
reconstruction of the original blade characteristic curves. This obviously leads to the convenient possibility of generating
parameter spaces having the desired dimension. In addition, a further relevant advantage of such parameterization
strategy based on splines control points displacement, is that all the blades generated are smooth deformations of the original one.
Figure \ref{bladex_PPTC} shows a pitch curve reconstruction by means of a 10 control points 3rd order spline carried out through BladeX.
In the example, non null displacements are also assigned to control points 6 and 7, to generate a modified pitch distribution, which would
ultimately result in a different blade geometry.

Once the parameterized blade geometry has been generated, the full propeller geometry can be finalized by replicating the blade for the
desired number of times, and attaching it to the imported hub geometry.

\subsection{A family of PPTC SVA-VP1304 blade deformations for the optimization campaign}

We based our analysis on the shape of the PPTC SVA-VP1304 benchmark propeller~\footnote{Geometry and documentation
available at \url{https://www.sva-potsdam.de/en/potsdam-propeller-test-case-pptc}}, originally designed for the
SMP workshops \cite{barkmannPPTC}.  To carry out the numerical experiments, we produced a set
of $1100$ blade variants, based on deforming the pitch and camber radial distributions along the blade. More specifically, the
deformations were obtained displacing the 10 control points of the splines reconstructing the pitch and camber profiles,
within $15 \%$ and $20 \%$ of the original blade maximum local pitch and maximum local camber, respectively. We point out that
the camber modification is carried out by scaling the camber line points of each sectional airfoil so as to obtain the specified
local maximum camber deflection. The described methodology resulted in a family of deformations depending on $20$ parameters.
As the blade profiles obtained from such procedure might suffer from inflections which might lead to unfeasible manufacturing
as well as the poor hydrodynamic performance, the local deformation bounds were also constrained in a way that ensures smooth
profiles.

\section{Parameter space analysis through active subspaces}
\label{sec:active}
In this section, we present the active subspaces analysis of the fluid dynamic performance results obtained for each  
PPTC SVA-VP1304 benchmark propeller variation produced. Such results were obtained using the potential flow solver
PROCAL \cite{vazPROCAL} to simulate the flow past the propeller in an open water test setup.

\begin{figure}[htp!]
\centering
\subfloat[][]{\includegraphics[width=.41\textwidth]{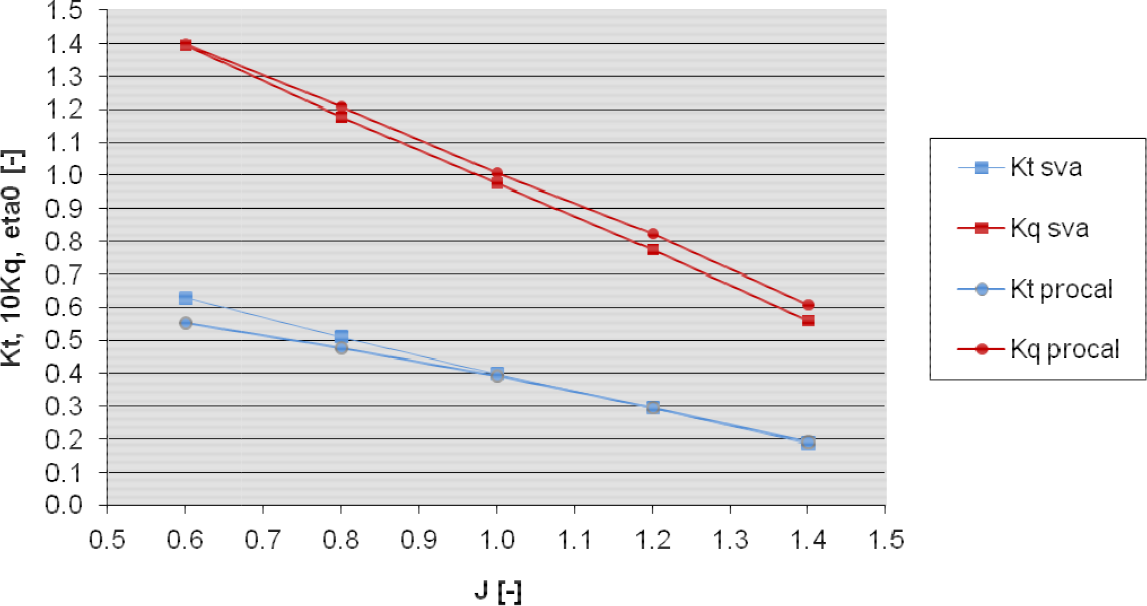}\label{fig:procal}}\hfill
\subfloat[][]{\includegraphics[width=.49\textwidth]{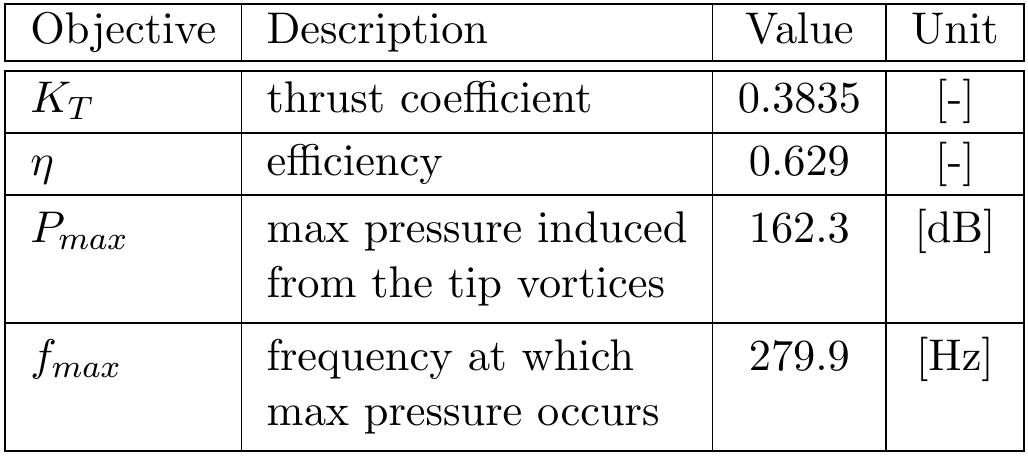}\label{fig:orig_charac}}
\caption{(a) PROCAL prediction of the thrust and torque coefficients of the PPTC SVA-VP1304 at various advance ratios. Results
are compared with the corresponding experimental data from the SMP workshop. (b) Computed output parameters at $J = 1.019$.}
\label{fig:example}
\end{figure}

The present study was carried out for a single value of the propeller advance ratio $J = \tfrac{V_a}{n \cdot D} = 1.019$ where
$V_a$ is the streamwise velocity, $D=0.25 m$ is the propeller diameter, and $n$ is the rotational speed in (rps). While $J$ is
a parameter summarizing the fluid dynamic inputs to the simulations, the first outputs of interest for the designers are quite
naturally an estimation of the  hydrodynamic forces and moments acting on the propeller. In particular, the thrust force $T$ generated
by the propeller along its axial direction is the quantity that the designers typically want to maximize. At the same time, the resisting
torque $Q$ around the propeller axis needs instead to be minimized to increase performance. Based on such considerations, the first
output parameter considered in this work is the thrust coefficient $K_T = \tfrac{T}{\rho n^2 D^4}$ ($\rho$ being the fluid density).
As for the second output parameter, we preferred using the propeller efficiency $\eta = \tfrac{J}{2 \pi} \cdot \tfrac{K_T}{K_Q}$
rather than simply using the torque coefficient $K_Q = \tfrac{Q}{\rho n^2 D^5}$. A high propeller efficiency is in fact a significant
indicator of the propeller ability to generate thrust, without requiring high torque values from the engine to mantain the
indicator of the propeller ability to generate thrust, without requiring high torque values from the engine to maintain the
specific rotational speed. For the value of $J$ herein considered, the efficiency and thrust coefficient obtained for the original
PPTC  SVA-VP1304 benchmark propeller are $\eta = 0.629$, $K_T = 0.3835$ respectively. As shown in Figure~\ref{fig:procal},
at the selected advance ratio the propeller thrust coefficient as predicted by a non-cavitating unsteady PROCAL computation is very
close to the experimental thrust coefficient (SMP'11 workshop; test case 2.3.1 \cite{barkmannPPTC}), the difference
amounting to less than $1 \%$. Along with the aforementioned propeller thrust coefficient $K_T$ and efficiency $\eta$, the
output parameters also considered in the analysis were the vortex-induced maximum pressure ($P_{max}$), and the
frequency ($f_{max}$) associated to ($P_{max}$). A summary of the values of the four outputs for the benchmark propeller are presented
in Figure~\ref{fig:orig_charac}.

The procedure adopted in the present study is composed as follows: (i) dimension reduction via the active subspaces analysis
on the geometrical parameter space defined by the the control points deformations, (ii) sensitivity analysis and optimization
of the propeller performance based on the reduced parameter obtained from the active subspaces analysis. In the following
subsections, we first provide a brief description of the active subspaces property theory, and then report the results of
the analysis carried out on the potential flow
results database.

\subsection{Background and formulations}

The active subspaces (AS) property has been recently establishing as one of the emerging techniques for dimension reduction in
parametric studies~\cite{constantine2015active,constantine2014active}. Since its introduction, AS has been widely applied in
several research topics, including marine engineering~\cite{tezzele2018dimension,tezzele2018model,demo2018isope}, and
cardio-vascular flows~\cite{tezzele2018combined}. A parameter study of an objective function $f (\mupar)$ becomes challenging
when the dimension of $\mupar$ (i.e.: the number of input parameters considered) is relatively large. In that regard
AS offer a sophisticated approach to reduce the study's dimensions by seeking a set of important directions in the parameter
space along which $f$ varies the most. Such directions are linear combinations of all the parameters, and span a lower
dimensional subspace of the input space, which can be also exploited to carry out optimization campaigns in an extremely
inexpensive fashion. 

Consider the objective $f (\mupar): \mathbb{D} \subset \mathbb{R}^m \rightarrow \mathbb{R}$ as a differentiable, square-integrable scalar function of the normalized inputs. In order to determine the directions of maximum variability we evaluate the uncentered covariance matrix of gradients $\mathbf{C} = \mathbb{E} [(\nabla_{\mupar} f) ( \nabla_{\mupar} f )^T] = \int_{\mathbb{D}} (\nabla_{\mupar} f) ( \nabla_{\mupar} f )^T \rho \, d \mupar$, where $\mathbb{E}[\cdot]$ is the expectation operator, and $\rho: \mathbb{D} \rightarrow \mathbb{R}^+$ is the probability density function. The symmetric positive definite (SPD) structure of $\mathbf{C}$ allows for an eigendecomposition, $\mathbf{C} = \mathbf{W} \bm{\Lambda} \mathbf{W}^T,$ where $\mathbf{W}$ is the $m \times m$ column matrix of eigenvectors, and $\bm{\Lambda}$ is the diagonal matrix of non-negative eigenvalues arranged in descending order. Now by partitioning $\bm{\Lambda} = \bigl[ \begin{smallmatrix}\Lambda_1 & \\ & \Lambda_2 \end{smallmatrix}\bigr]$ into the larger eigenvalues, $\Lambda_1 = \textrm{diag}\{\lambda_1, \dots, \lambda_M\}$, and the smaller ones, $\Lambda_2 = \textrm{diag}\{\lambda_{M+1}, \dots, \lambda_m\}$, subsequently $\mathbf{W} = [\mathbf{W}_1 \quad \mathbf{W}_2]$ such that $\mathbf{W}_1 \in \mathbb{R}^{m \times M}$, $\mathbf{W}_2 \in \mathbb{R}^{m \times m-M}$, then the low eigenvalues $\Lambda_2$ suggest that the corresponding vectors $\mathbf{W}_2$ are in the null space of the covariance matrix $\mathbf{C}$, and such vectors can be discarded to form an approximation. Therefore the lower dimensional parameter subspace spanned by $\mathbf{W}_1$ is considered as the active subspace, while the inactive subspace is spanned by $\mathbf{W}_2$. At this stage, the dimension reduction is achieved by projecting $\mupar$ onto the active subspace to obtain the active variables $\mupar_M = \mathbf{W}_1^T\mupar \in \mathbb{R}^M$, whereas the inactive variables are $\zetapar = \mathbf{W}_2^T \mupar \in \mathbb{R}^{m - M}$. The relationship between the full parameter space $\mupar \in \mathbb{D}$ and the active variables $\mupar_M$ is described as $\mupar = \mathbf{W}_1\mathbf{W}_1^T\mupar + \mathbf{W}_2\mathbf{W}_2^T\mupar = \mathbf{W}_1 \mupar_M + \mathbf{W}_2 \zetapar$, and the objective function $f (\mupar)$ is approximated by $g (\mupar_M)$ which can be further exploited to construct a response surface.


\subsection{Sensitivity analysis and optimization using active subspaces}

\begin{figure}[htp!]
\centering
\subfloat[][$K_T$]{\includegraphics[width=.24\textwidth]{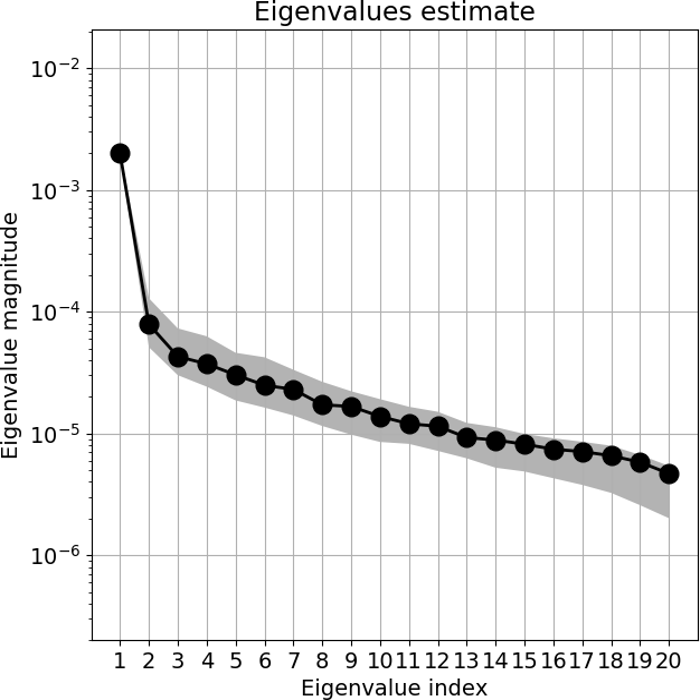}\label{fig:val_Kt}}
\subfloat[][$\eta$]{\includegraphics[width=.24\textwidth]{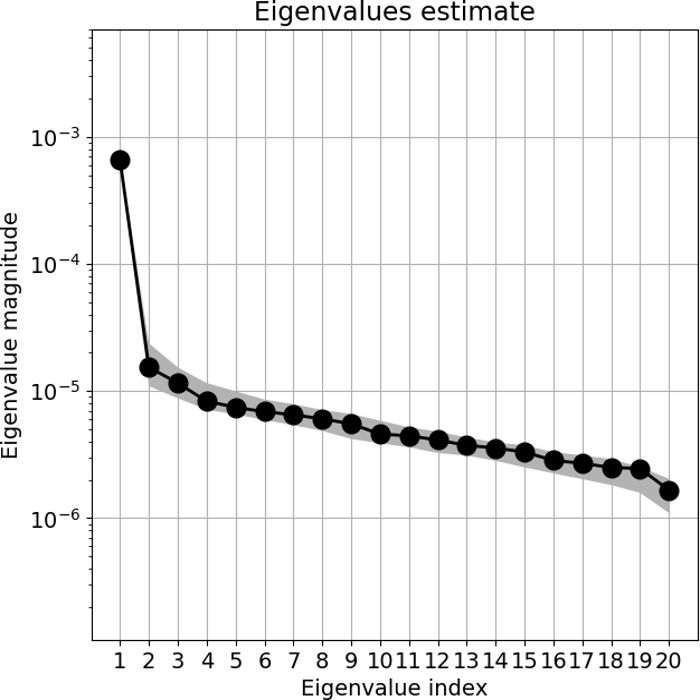}\label{fig:val_eta}}
\subfloat[][$P_{max}$]{\includegraphics[width=.24\textwidth]{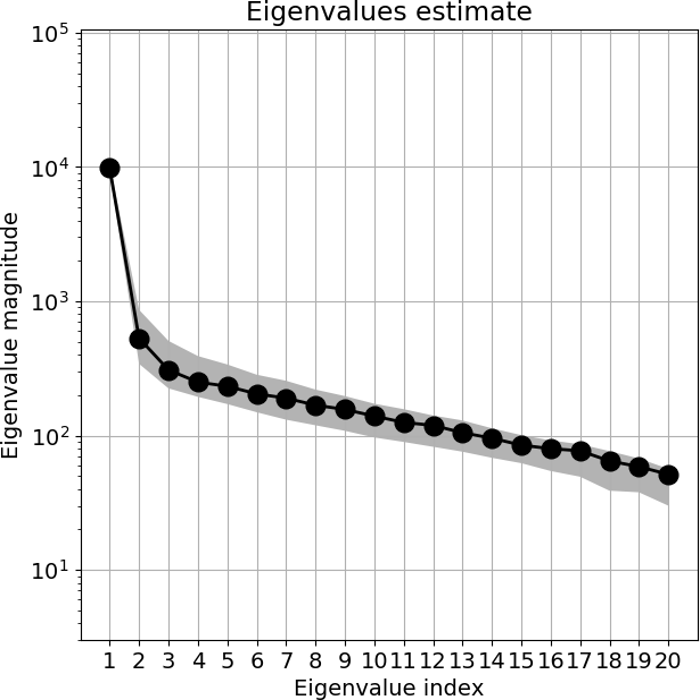}\label{fig:val_pm}}
\subfloat[][$f_{max}$]{\includegraphics[width=.24\textwidth]{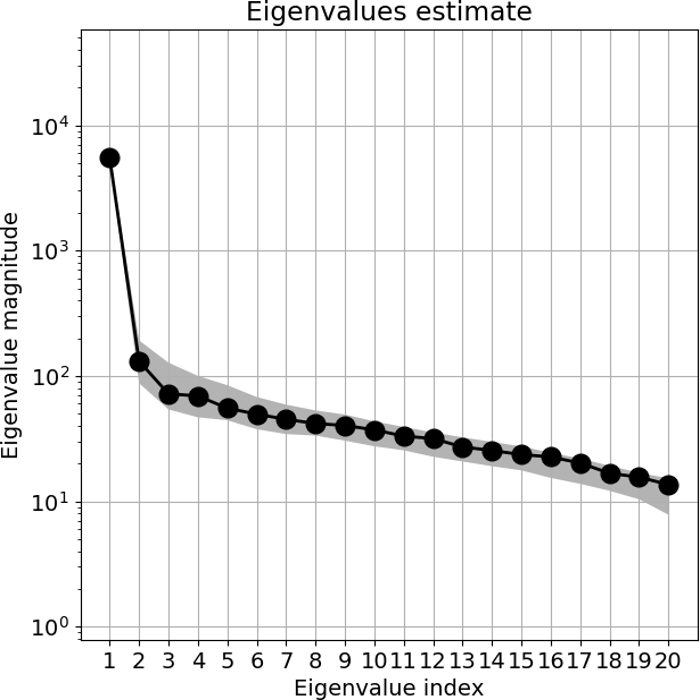}\label{fig:val_f}}
\caption{Eigenvalues of the uncentered covariance matrix of gradients, relating the geometrical parameters $\mupar \in \mathbb{R}^{1100 \times 20}$ to each of $K_T$, $\eta$, $P_{max}$, or $f_{max}$ represented by $f(\mupar) \in \mathbb{R}^{1100}$. The low eigenvalues suggest the corresponding eigenvectors are in the null space of the covariance matrix, and thus a one dimensional active variable can be achieved as an approximation of $\mupar$.}
\label{fig:eigvals}
\end{figure}

\begin{figure}[htp!]
\centering
\subfloat[][$K_T$]{\includegraphics[width=.24\textwidth]{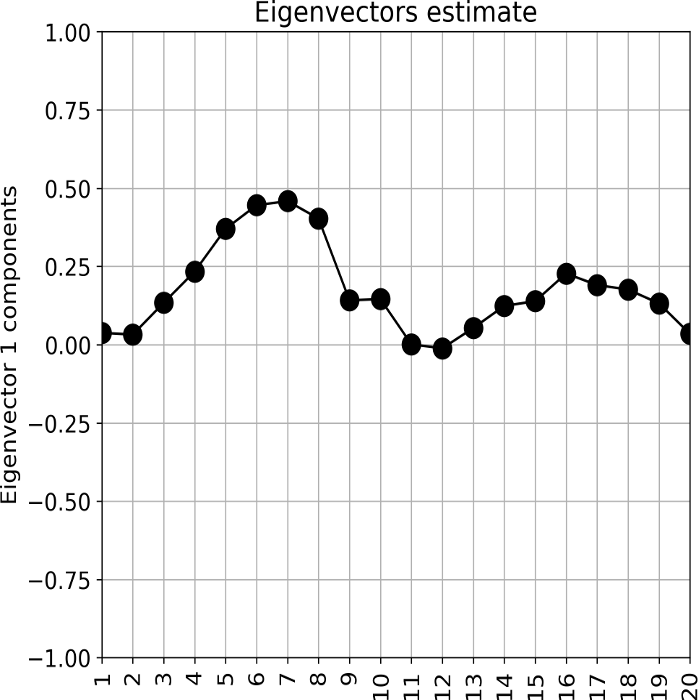}\label{fig:vec_Kt}}
\subfloat[][$\eta$]{\includegraphics[width=.24\textwidth]{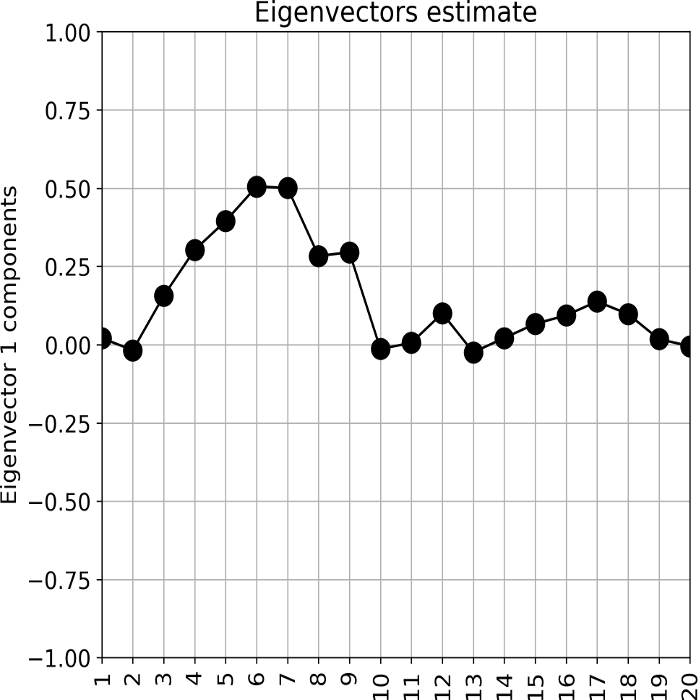}\label{fig:vec_eta}}
\subfloat[][$P_{max}$]{\includegraphics[width=.24\textwidth]{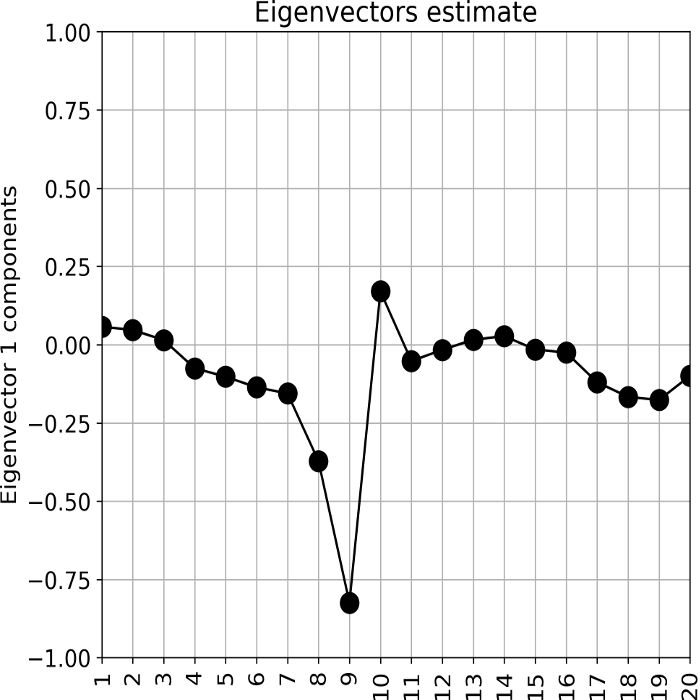}\label{fig:vec_pm}}
\subfloat[][$f_{max}$]{\includegraphics[width=.24\textwidth]{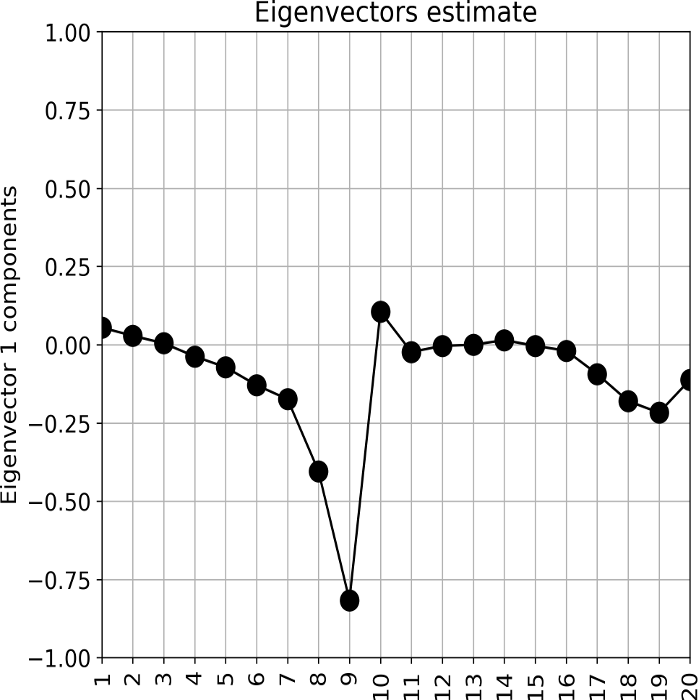}\label{fig:vec_f}}
\caption{Components of the first eigenvector, i.e. the active subspace $\mathbf{W}_1$, which describes the contribution of each of the $20$ parameters in the AS approximation. The plots show that $K_T$ and $\eta$ are mostly sensitive to the mid-to-near-tip region of the pitch profile, while $P_{max}$ and $f_{max}$ are mostly sensitive to the near-tip region of the pitch curve.}
\label{fig:eigvecs}
\end{figure}

According the AS formulation presented, we consider the geometrical parameters
$\mupar \in \mathbb{R}^{1100 \times 20}$ which represent the displacements of the $20$ control points of all
the $1100$ shapes. As for the
parameters ordering in vector $\mupar$, the first 10 parameters represent the pitch spline control point displacements,
going from the blade root to the tip. The last 10 parameters are the camber line spline control point displacements,
again ordered from root to tip. The objective function is $f_i(\mupar) \in \mathbb{R}^{1100}$, where the
index $i=1,\dots,4$ indicates the specific output parameter considered, in the order $K_T$, $\eta$, $P_{max}$, or $f_{max}$. The
eigendecomposition was performed on the covariance matrices corresponding to each output parameter and the resulting eigenvalues
magnitudes are presented in Figure~\ref{fig:eigvals}. The plots clearly show that for all the output parameter considered, a
significant gap exists between the magnitude of first eigenvalue and that of the remaining eigenvalues. This observation
suggests that each of the the input to output relationships can be rather accurately represented with a one dimensional approximation.
Such one dimensional relationship is computed as the projection of
the parameter space $\mupar$ onto the active subspace corresponding to the first eigenvalue (i.e.: the first eigenvector), namely
$\mupar_M = \mupar \cdot \mathbf{W}_1 \in \mathbb{R}^{1100}$. In Figure~\ref{fig:eigvecs} we show present an attempt to visualize the
subspace $\mathbf{W}_1 \in \mathbb{R}^{20}$. The $20$ components in each plots represent in fact the weights
needed to obtain the active variable as a linear combination of the of the original input parameters.
So, such visualization is able to indicate which parameters have a higher influence on the output, as the corresponding components will be
characterized by higher weight magnitudes. The results suggest that both $K_T$ and $\eta$ are mostly sensitive to the mid-to-near-tip region
of the pitch profile, whereas the $P_{max}$ and $f_{max}$ are mostly sensitive to the near-tip region of the pitch curve. In
fact, the resulting sensitivity analysis coincides with the hydrodynamic experience and the design practice, where the pitch is
directly related to the loading on the propeller and to the tip vortex strength. In addition, the efficiency is directly proportional
to the thrust by definition, and the blade loading, thus the $K_T$, is much affected by the pitch at the radial coordinate range
around $0.7 r/R$. Such radial coordinate is in fact used in common propeller descriptions, to provide a meaningful reference value for
pitch and loading. Moreover, the plots suggest that the pitch at the tip has the largest impact on the tip vortex pressure $P_{max}$ and subsequentially $f_{max}$. As
for the camber modifications, they appear to have on the loading a lower but still significant impact  with respect to the pitch deformations,
and an even less relevant effect on the tip vortex strength. A complete summary of the parameters influence on the propeller
performance is presented in Table~\ref{tab:summary}.

\begin{table}[htb!]
\centering
\caption{Summary of the PPTC performance sensitivity towards the pitch and camber root-tip parametric curves. In the table, ($++$) represents a dominating influence, ($+$): significant influence, ($+-$): small influence, and ($-$): can be neglected.} \label{tab:summary}
\begin{tabular}{|l|c|c|c|c||l|c|c|c|c|}
\hline
\hline
Control points & $\bm k_t$ & $\bm \eta$ & $\bm p_m$ & $\bm f_m$ & Control points & $\bm k_t$ & $\bm \eta$ & $\bm p_m$ & $\bm f_m$ \\ \hline
pitch - 1 & -- & -- & +-- & +-- & camber - 1 & -- & -- & +-- & -- \\ \hline
pitch - 2 & -- & -- & +-- & -- & camber - 2 & -- & +-- & -- & -- \\ \hline
pitch - 3 & + & + & -- & -- & camber - 3 & -- & -- & -- & -- \\ \hline
pitch - 4 & + & + & +-- & -- & camber - 4 & + & -- & -- & -- \\ \hline
pitch - 5 & ++ & + & + & +-- & camber - 5 & + & +-- & -- & -- \\ \hline
pitch - 6 & ++ & ++ & + & + & camber - 6 & + & +-- & -- & -- \\ \hline
pitch - 7 & ++ & ++ & + & + & camber - 7 & + & +-- & + & +-- \\ \hline
pitch - 8 & ++ & + & ++ & ++ & camber - 8 & + & +-- & + & + \\ \hline
pitch - 9 & + & + & \textbf{++} & \textbf{++} & camber - 9 & + & -- & + & + \\ \hline
pitch - 10 & + & -- & ++ & + & camber - 10 & -- & -- & + & + \\ \hline
\bottomrule
\end{tabular}
\end{table}

\begin{figure}[htp!]
\centering
\subfloat[][]{\includegraphics[width=.32\textwidth]{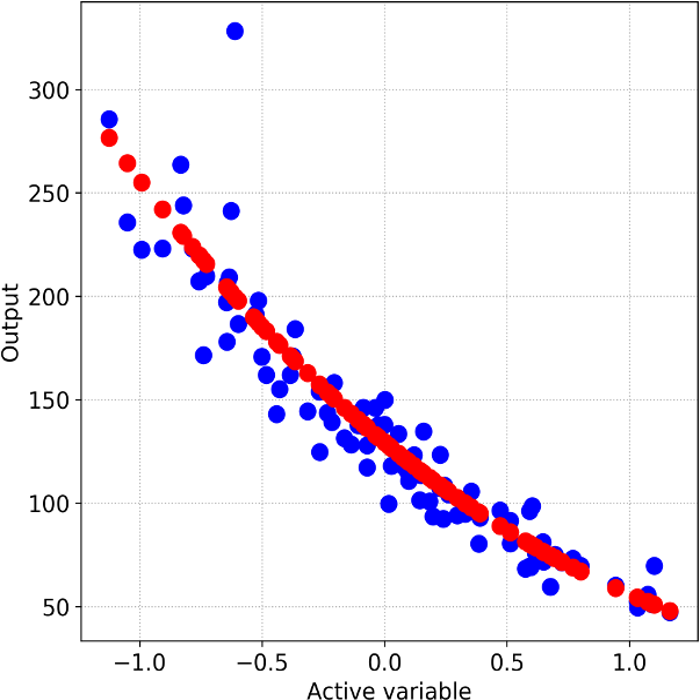}\label{fig:RS}} \hspace{25mm}
\subfloat[][]{\includegraphics[width=.3\textwidth]{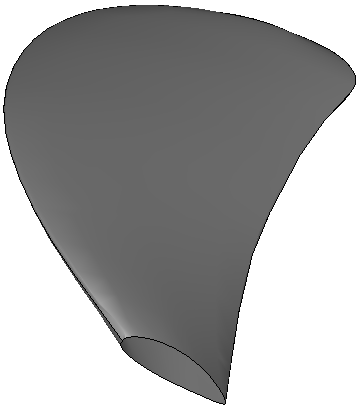}\label{fig:min_blade}}
\caption{(a) Response surface (RS) of the reduced parameter $\mupar_M$ vs. $P_{max}$ constructed as a best-fit polynomial trained from $80\%$ of the dataset, the remaining $20\%$ are used to validate the output (in blue) and the corresponding RS (in red). (b) The morphed PPTC blade to produce a minimal $P_{max}$.}
\label{fig:minimization}
\end{figure}

\begin{figure}[htp!]
\centering
\subfloat[][]{\includegraphics[trim=30 0 30 30, width=.49\textwidth]{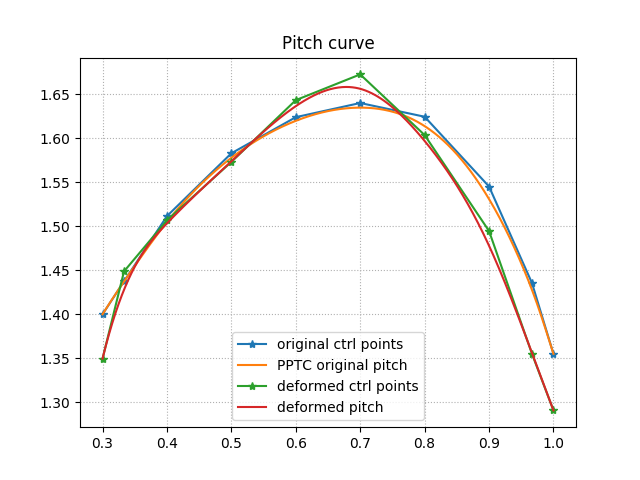}} \hfill
\subfloat[][]{\includegraphics[trim=30 0 30 30, width=.49\textwidth]{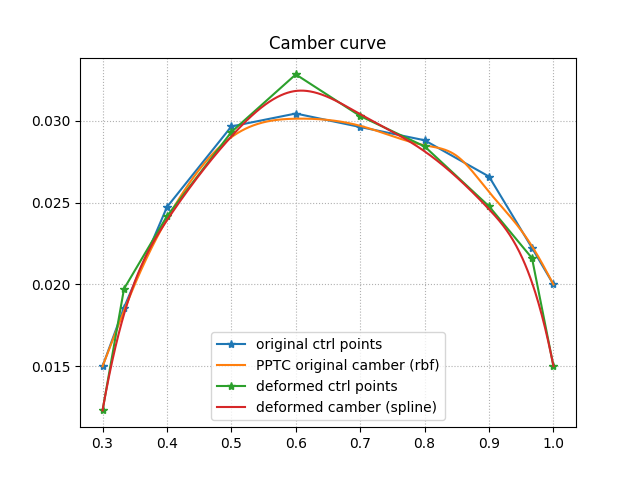}}
\caption{Deformed parametric curves resulting from the minimization procedure for $P_{max}$. (a) Pitch, (b) camber.}
\label{fig:min_profiles}
\end{figure}

We now describe a further possibility offered by AS analysis. We in fact exploit the input to output
relationship in the active subspace to carry out an optimization campaign in a low dimensional ---hence reduced--- space.
For instance, if we consider the tip vortex-induced pressure $P_{max}$, we can readily represent its dependence on its
active variable $\mupar_M$ with a one dimensional response surface, as depicted in Figure~\ref{fig:RS}. Such response
surface is then conveniently used to
determine the active variable corresponding to the minimal $P_{max}$. The resulting optimal $\mupar_M$ value is then mapped back to the
actual parameter space so as to identify the exact root-tip deformations
yielding the minimal acoustic pressure, as reported in Figures~\ref{fig:min_blade} and~\ref{fig:min_profiles}. The deformed profiles
were utilized via BladeX to
construct the morphed blade, Figure~\ref{fig:min_blade}. Finally, since the ultimate goal was to minimize $P_{max}$ and $f_{max}$, 
maximize $\eta$ without altering $K_T$, such procedure had to be implemented by introducing a shared subspace~\cite{ji2018shared}
among the four objectives, and a constrained optimization needed to be carried out on the resulting response surface in order to
find the optimal propeller. Among the $1100$ variants produced, the shape resulting from the procedure described was eventually
selected to undergo a high fidelity RANS simulations.

\section{Conclusions and perspectives}
\label{sec:the_end}
In the present contribution, we presented an application of parameter space reduction based on the Active Subspaces (AS) property, in
the framework of the hydroacoustic optimization of ship propellers. Making use of the open source Python package BladeX,
we produced a large number of parameterized modifications of the PPTC SVA-VP1304 benchmark propeller, which were used to carry out
potential flow simulations with the software PROCAL. The AS analysis suggested that for all the propeller performance parameters
considered the input to output relationships can be rather accurately represented with a one dimensional approximation, in which
the single active parameter is a linear combination of the 20 original shape parameters. A further sensitivity analysis based on
the weights of such linear combination suggested that, at a first approximation, the pitch modifications in the mid-to-tip region
and --- at a lesser extent --- the camber modification in the blade middle portion have highers impact on the output. 

These results open interesting perspective on the application of parameter space reduction in naval engineering problems. Possible
developments could be obtained by testing the possibility of carrying out similar investigations employing reduced fluid dynamic 
models, as the ones broadly described in~\cite{morhandbook2019} and~\cite{rozza2018advances}. In particular, the use of reduced order
models based on POD would allow for fluid dynamic simulations that account for all the relevant physical phenomena in the flow,
at a computational cost compatible with the present analysis. Ongoing work in the PRELICA project is directed in such direction.

\section*{Acknowledgements}
This work was partially performed in the context of the PRELICA -
``Advanced methodologies for hydro-acoustic design of naval
propulsion'', supported by Regione
FVG, POR-FESR 2014-2020, Piano Operativo Regionale Fondo Europeo per
lo Sviluppo Regionale, and partially supported by European Union Funding for
Research and Innovation --- Horizon 2020 Program --- in the framework
of European Research Council Executive Agency: H2020 ERC CoG 2015
AROMA-CFD project 681447 ``Advanced Reduced Order Methods with
Applications in Computational Fluid Dynamics'' P.I. Gianluigi
Rozza.

\end{document}